\documentclass[12pt]{cernrep}
\usepackage{graphicx,epsfig,epsf,psfig,amssymb,
rotating,colordvi,helvet,color,subfigure}
\def\s#1{{\small#1}}

\def\HW{\s{HERWIG}}

\def\B0bar{\overline{B^0}}

\begin{document}
 \title{CHARGED HIGGS BOSONS IN THE TRANSITION REGION 
\boldmath${M_{H^\pm} \sim m_t}$
 AT THE LHC}
\author{K.A. Assamagan$^1$, M. Guchait$^2$ and S. Moretti$^3$}
\institute{$^1$BNL, Upton NY, USA; $^2$TIFR, Mumbai, India; $^3$Southampton 
University, UK}
\maketitle
\begin{abstract}
We illustrate preliminary results obtained through Monte Carlo ({\small HERWIG}) and
detector ({\small ATLFAST}) simulations of the $H^\pm\to\tau^\pm\nu_\tau$ 
signature of charged Higgs bosons with masses comparable to that of the
top quark.
\end{abstract}

\section{THE THRESHOLD REGION}

The detection of charged Higgs bosons ($H^\pm$) would
unequivocally imply the existence of physics beyond the Standard Model (SM),
since spin-less charged scalar states do not belong to its particle spectrum.
Singly charged Higgs bosons appear in any Two-Higgs Doublet Model (2HDM),
including a Type-II  in presence of minimal Supersymmetry (SUSY), namely, 
the Minimal Supersymmetric Standard Model (MSSM). Depending on its mass,
the machines that
are likely to first discover such a state are Tevatron 
$p\bar p$ ($\sqrt s=2$ TeV) and the LHC ($\sqrt s=14$ TeV). Current limits 
on the charged Higgs boson mass are set by LEP at about 
80~GeV. At the Tevatron a charged Higgs boson could be 
discovered for masses up to $m_t-m_b$, whereas the LHC
has a reach up to the TeV scale, if $\tan\beta$ is favourable
(i.e., either large or small).

\begin{figure}[!h]
\vspace{-0.25cm}
\begin{center}
\epsfig{file=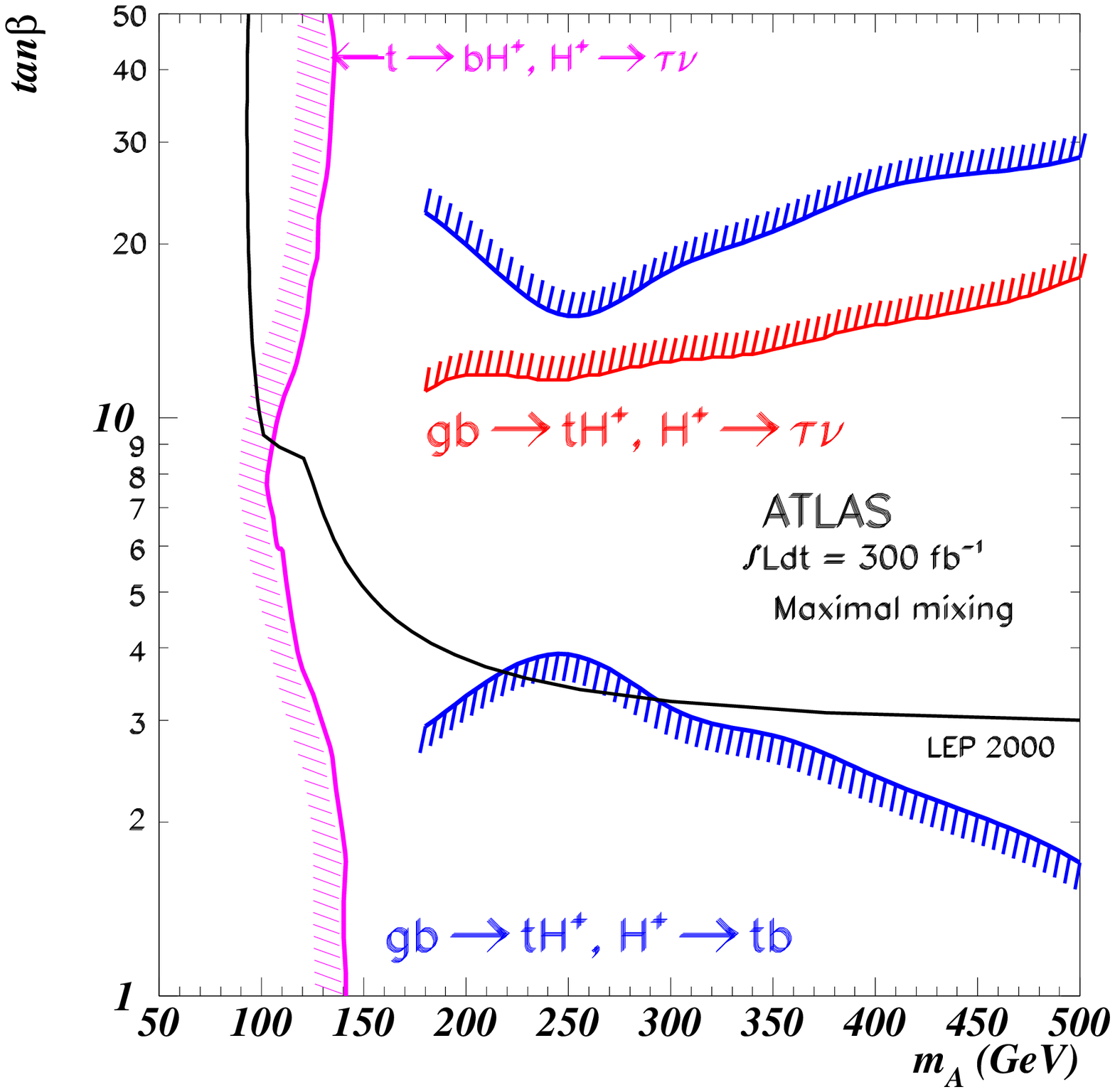,width=6.5cm,height=5.0cm}
\epsfig{file=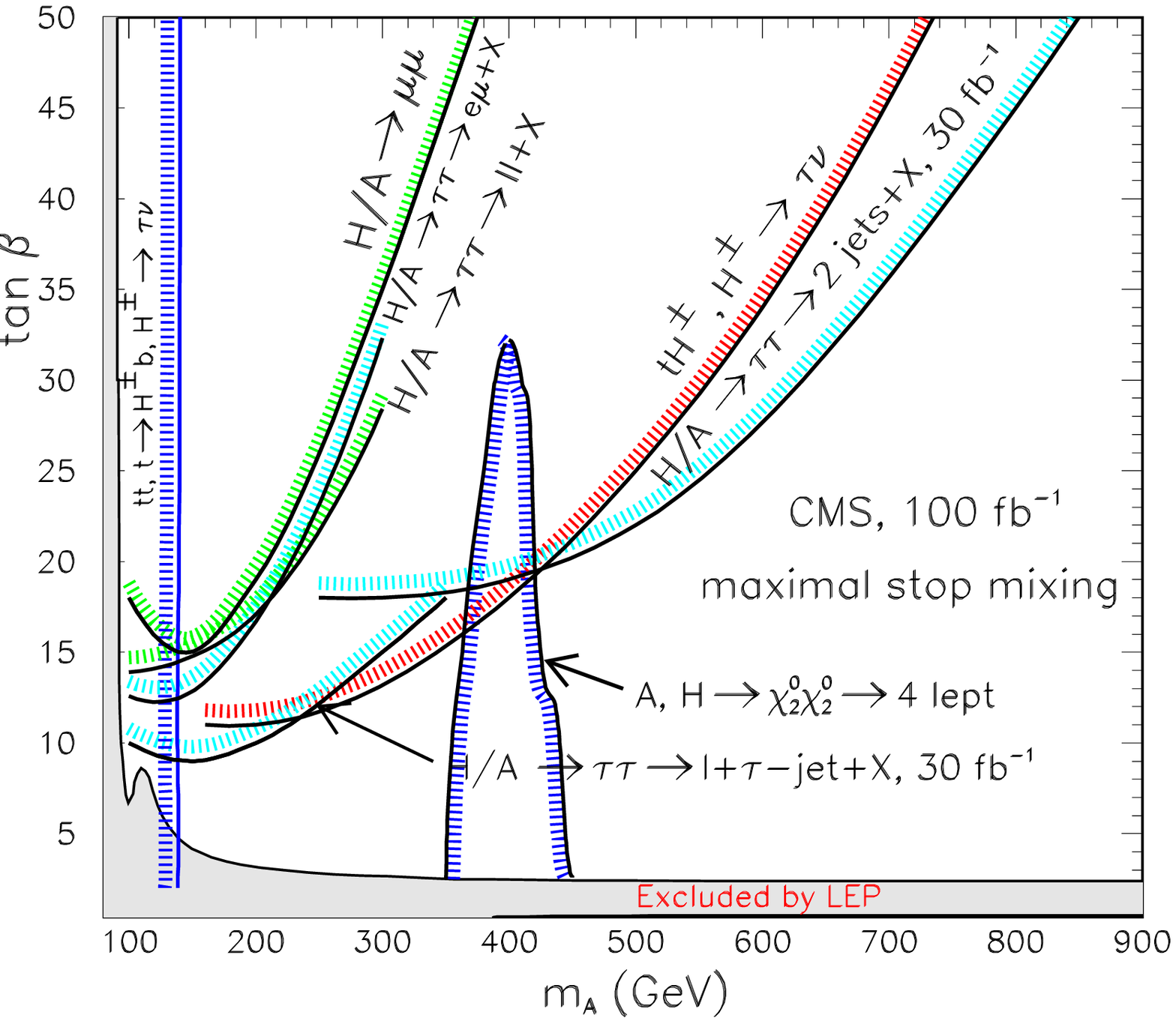,width=6.0cm,height=5.0cm}
\caption{\small The ATLAS 5-$\sigma$ discovery contours of 
2HDM charged Higgs 
bosons for 
300~fb$^{-1}$ of luminosity, only including the reach of SM decay modes  (left plot).
The CMS 5-$\sigma$ discovery contours of MSSM Higgs bosons for 100~fb$^{-1}$ of 
luminosity, also including the reach of $H,A \to \chi^0_2 \chi^0_2 \to 4l^{\pm}$ decays, 
assuming $M_1$ = 90~GeV, $M_2$ = 180~GeV, $\mu$ = 500~GeV, $M_{\tilde{\ell}}$ = 250~GeV, 
$M_{\tilde{q},\tilde{g}}$ = 1000~GeV (right plot). }
\label{fig:LHC}
%{\vskip-6.6cm\hskip-6.0cm{\tiny{\Blue{$gb\to tH^+$, $H^+\to t\bar b$}}}}
%\vskip+5.6cm
\end{center}
\end{figure}

\begin{figure}[!h]
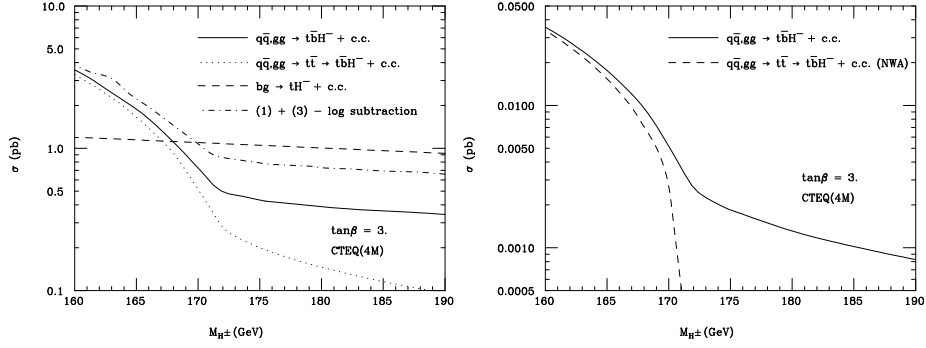

%\vspace{-0.25cm}
\begin{center}
\epsfig{file=LHCthreshold3_LH.eps,angle=90,height=4.5cm}
\epsfig{file=TEVthreshold3_LH.ps,angle=90,height=4.5cm}
\caption{\small  
Cross section for $gg,q\bar q\to t\bar b H^-$,
$gg,q\bar q \to t\bar t \to t\bar b H^-$ with finite top quark width, 
$bg\to tH^-$ and
the combination of the first and the last, at the LHC with $\sqrt s=14$ TeV
(left plot). Cross section for $gg,q\bar q\to t\bar b H^-$ and
$gg,q\bar q \to t\bar t \to t\bar b H^-$ in NWA, at 
the Tevatron with $\sqrt s=2$ TeV (right plot). Rates are
function of $M_{H^\pm}$ for a representative value of $\tan\beta$.}
\label{fig:threshold}
\vspace{-0.25cm}
\end{center}
\end{figure}

For the LHC, 
the ATLAS discovery potential of $H^\pm$ bosons in a general Type-II 2HDM  
or MSSM 
(prior to the results of this study) is visualised in the left-hand side
of Fig.~\ref{fig:LHC}. (A similar CMS plot, also including neutral Higgs states,
is given for comparison.)
The existence of a gap in coverage for $M_{H^\pm}\approx m_t$ was already
denounced in Refs.~\cite{Cavalli:2002vs,Moretti:2002ht} as being due to
the fact that Monte Carlo (MC) simulations of $H^\pm$ production for
$M_{H^\pm}\sim m_t$ were flawed by a wrong choice of the hard scattering
process. In fact, for $M_{H^\pm}< m_t$, the  estimates in both plots in
Fig.~\ref{fig:LHC}
were made by assuming 
as  main production mode of $H^\pm$ scalars the decay 
of top
(anti)quarks produced via QCD in the annihilation
of gluon-gluon and quark-antiquark pairs (hence -- by definition -- 
the attainable Higgs mass is strictly confined to 
the region $M_{H^\pm}\le m_t-m_b$).
This should not be surprising (the problem was also
encountered by CMS, see right-hand side of
Fig.~\ref{fig:LHC}), since standard MC programs, such as
{\small PYTHIA} and \HW\ \cite{Sjostrand:2003wg,Corcella:2000bw}, have 
historically accounted for this process through the usual procedure of 
factorising
the production mode, $gg,q\bar q\to t\bar t$, times the
decay one, $\bar t\to \bar b H^-$, in the so-called Narrow Width
Approximation (NWA) \cite{Guchait:2001pi}. This description
fails to correctly account for the production 
phenomenology of charged Higgs bosons when their mass approaches or indeed
exceeds that of the top-quark (i.e., falls in the so called `threshold 
region'). This is evident from the left plot in 
Fig.~\ref{fig:threshold}. (The problem also occurs at
Tevatron, see right plot therein
and Refs.~\cite{Guchait:2001pi,Alwall:2003tc}.) As remarked in
Ref.~\cite{Guchait:2001pi}, the use of the $2\to 3$ hard scattering
process $gg,q\bar q\to t\bar b H^-$  
\cite{Gunion:1994sv}--\cite{Belyaev:2002eq}, in place of the
`factorisation' procedure in NWA, is mandatory in the threshold
region, as the former correctly keeps into account 
both effects of the finite width of the top quark
and the presence of other $H^\pm$ production mechanisms,
such as Higgs-strahlung and $b\bar t \to H^-$ fusion (and relative 
interferences). 
The differences seen between the two descriptions in 
Fig.~\ref{fig:threshold} are independent of $\tan\beta$ and 
also survive in, e.g., $p_T$ and $\eta$ spectra \cite{Guchait:2001pi}.

One more 
remark is in order, concerning the LHC plot in Fig.~\ref{fig:threshold}.
In fact, at the CERN hadron collider, the above $2\to3$ reaction is
dominated by the $gg$-initiated subprocesses, rather than by 
$q\bar q$-annihilation, 
as is the case at the Tevatron. This means that a potential 
problem of double counting arises in the simulation of $t H^- X$ + c.c. 
events at the LHC, if one considers that Higgs-strahlung can also be 
emulated through the $2\to2$ process $bg\to t H^-$ + c.c., 
as was done in assessing the ATLAS (and CMS) discovery reaches 
in the $H^+\to t\bar b$ and $H^+\to \tau^+\nu_\tau$ 
channels for $M_{H^\pm}> m_t$ (see 
Refs.~\cite{Assamagan:2002ne,Denegri:2001pn}
for reviews). The difference between the
two approaches is well understood, and prescriptions exist for
combining the two, either through the subtraction of a common logarithmic
term \cite{Borzumati:1999th,Moretti:1999bw}
or by means of a cut in phase space \cite{Belyaev:2002eq}. 

If one then
looks at the most promising (and cleanest) charged Higgs boson decay 
channel, i.e., $H^\pm\to\tau^\pm\nu_\tau$~\cite{Moretti:1995ds}, while 
using the $gg,q\bar q\to t\bar bH^-$ + c.c. 
description and reconstructing the 
accompanying top quark hadronically, the prospects of $H^\pm$ detection 
should improve significantly for $M_{H^\pm}$ values close to $m_t$, eventually
leading to the closure of the mentioned gap. The $2\to3$ 
description of the $H^\pm$ 
production dynamics (as well as the spin correlations in $\tau$-decays
usually exploited in the ATLAS  $H^\pm\to\tau^\pm\nu_\tau$ analysis) have
been made available in version 6.4 \cite{Corcella:2001wc} of 
the \HW\  event generator (the latter also 
through an interface to {\small TAUOLA} \cite{Jadach:1990mz}), 
so that detailed
simulations of $H^\pm$ signatures
at both the Tevatron and the LHC are now possible
for the threshold region, including fragmentation/hadronisation and detector
effects. In the next section we will discuss
the details of an ATLAS analysis based on such tools that has lead 
to the closure of the mentioned gap through the discussed charged Higgs
decay channel. This analysis was initiated 
in the context of the 2003 Les Houches workshop.

\section{ANALYSIS}
\label{sec:ana}

The signal $gg \to tbH^\pm \to jjbb\tau\nu$ and the major backgrounds, 
$gg \to t\bar{t} \to jjb\tau\nu b$ and $q\bar q,qg,\bar qg \to W+\mbox{jets}$, are generated 
with {\small HERWIG} v6.4 in the default implementation 
except for
{\small CTEQ5L}~\cite{PDFs} Parton Distribution Functions
(PDFs). The detector is
simulated with {\small ATLFAST}~\cite{ATLFAST}. The 
{\small TAUOLA} package~\cite{Jadach:1990mz} is used for the polarisation of the $\tau$-lepton. 
The selection of the final state requires a multi-jet trigger with a $ \tau$-trigger:
\begin{description}
  \item[(1)~~] We search for one hadronic $\tau$-jet, two $b$-tagged jets and at least two 
light-jets, all with $p_T > 30$~GeV. Furthermore, the $\tau$-jet and the $b$-tagged jets are required
to be within the tracking range of the ATLAS Inner Detector, $|\eta| < 2.5$. We assume a 
$\tau$-tagging efficiency of 30\% and a $b$-tagging efficiency of 60\%(50\%) at low(high) 
luminosity. The efficiency of this selection is at the level of 1.31\% for the signal (e.g., 
at $M_{H^\pm}=170$~GeV), 1.25\% for $gg \to t\bar{t} \to jjb\tau\nu b$ events and 
$(0.36\,\times\,10^{-3})$\% for $W^\pm$+jets events.
   \item[(2)~~] We reconstruct the invariant masses of pairs of light-jets, $m_{jj}$, and keep 
those consistent with the $W^\pm$ mass: $|m_{jj}-M_W| < 25$~GeV. The associated 
top-quark is then reconstructed requiring $|m_{jjb}-m_t| < 25$~GeV. For the signal with a 
charged Higgs mass of 170~GeV, 0.68\% of signal events pass this selection criteria compared to 
0.73\% and $(0.45\,\times\,10^{-6})$\% for the $t\bar{t}$ and $W^\pm$+jets backgrounds, respectively. 
   \item[(3)~~] We require that the transverse momentum of the $\tau$-jet be greater than 100~GeV, the 
transverse missing momentum be greater than 100~GeV and the azimuthal opening angle 
between the $\tau$-jet and the missing momentum vector be greater than one radian. 
Indeed, in the signal, the $\tau$-lepton originates from a scalar particle ($H^\pm$) 
whereas in the background the $\tau$-lepton comes from the decay of a vector particle ($W^\pm$). 
This difference reflects in the polarisation state of the $\tau$ and leads to harder $\tau$-jets 
in the signal compared to the backgrounds \cite{Assamagan:2002ne}--\cite{Roy:1999xw}. 
Furthermore, to satisfy 
the large cut on the transverse missing momentum and because the charged Higgs is heavier 
than the $W^\pm$-boson, a much larger boost is required from the $W^\pm$- in the background than 
from the $H^\pm$-boson in the signal. As a result, the spectra of the 
azimuthal opening angle between the 
$\tau$-jet and the missing transverse momentum are different for signals
and backgrounds, as shown in 
Fig.~\ref{fig:angle_mT} (left plot). 

\begin{figure}[!htbp]
\begin{center}
\epsfig{file=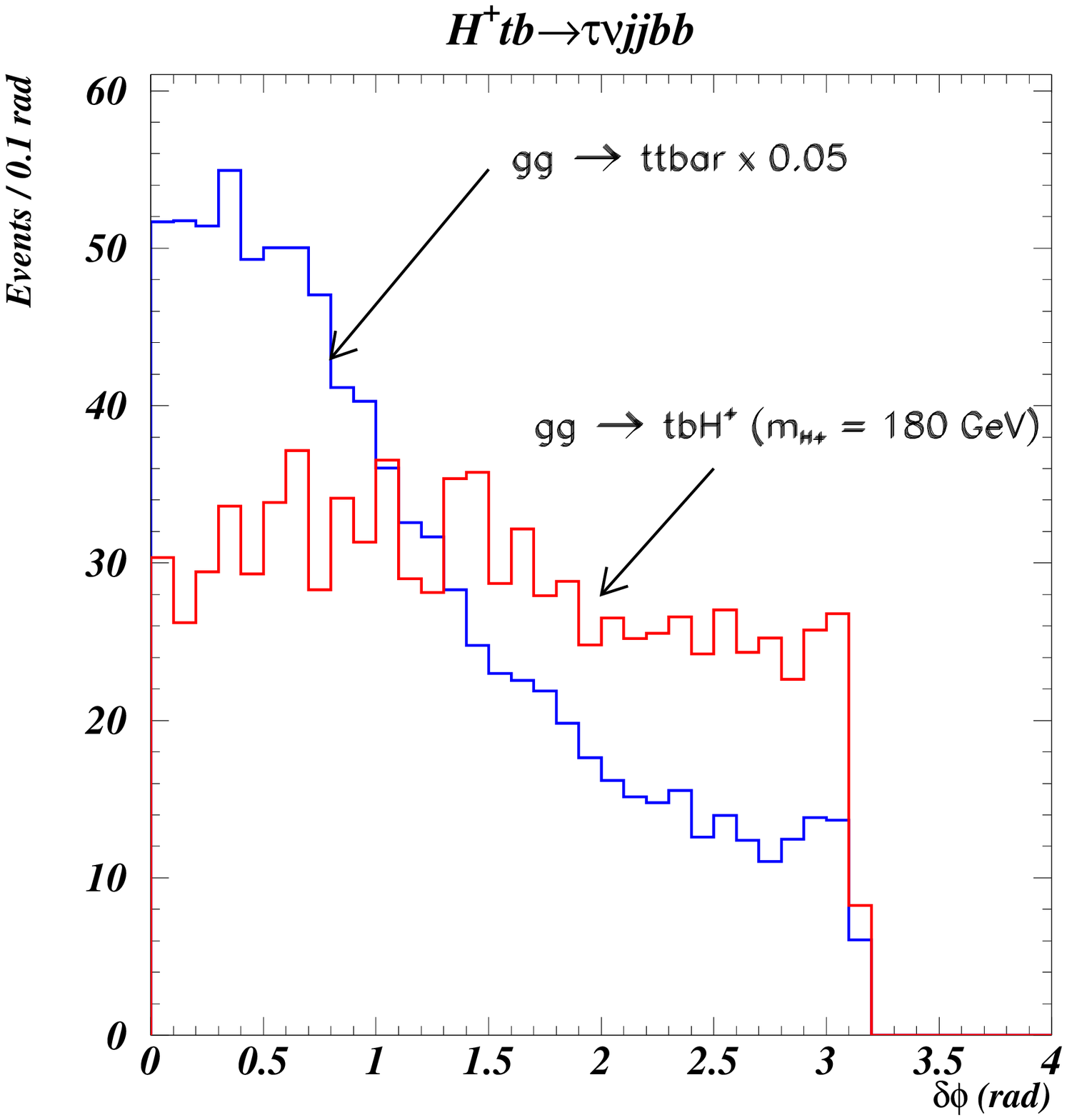,height=4.5cm}
\epsfig{file=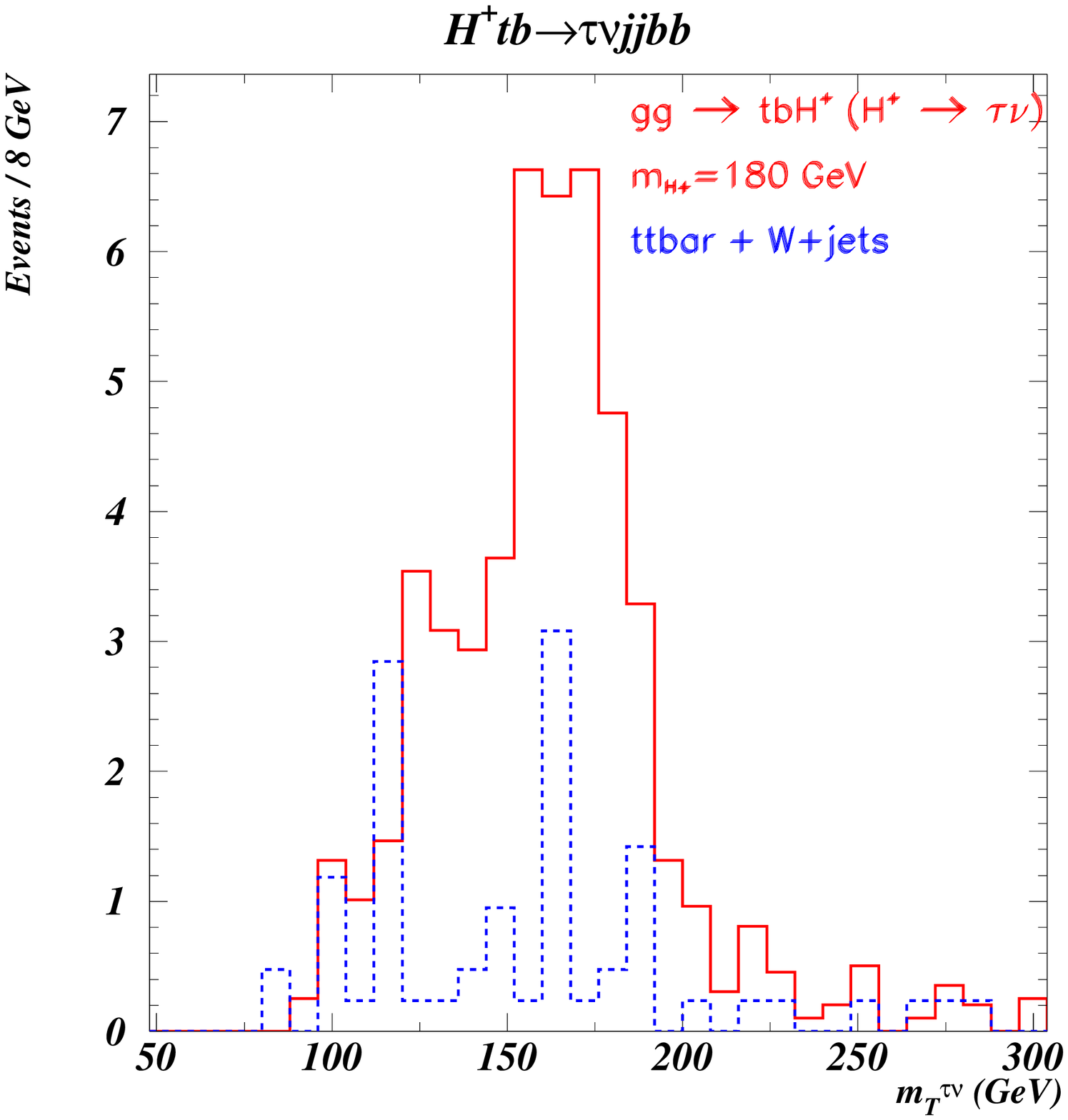,height=4.5cm}
\caption{\small The plot on the left shows the azimuthal opening angle between the $\tau$-jet 
and the transverse missing momentum. It peaks forward in the background and 
more and more backward in the 
signal, as the charged Higgs mass increases. The 
right plot shows the reconstructed transverse mass for a 180~GeV Higgs. (Both plots are shown 
for an integrated luminosity of 30~fb$^{-1}$.)}
\label{fig:angle_mT}
\vspace{-0.25cm}
\end{center}
\end{figure}

Although the full invariant mass of the $H^\pm \to \tau\nu$ system cannot be 
reconstructed because of the neutrino in the final state, the transverse mass (which is 
kinematically constrained to be below the $W^\pm$-mass in the backgrounds and below the 
$H^\pm$-mass in the signal)
\begin{equation}
\label{eq:trans}
m_T = \sqrt{2p_T^{\tau-{\rm jet}}{p\!\!\!/}_T\left[1-\cos(\Delta\phi)\right]}
\end{equation}
combines the benefits of both the polarisation effects and the kinematic boost, thus providing a 
good discriminating observable, as shown in Fig.~\ref{fig:angle_mT} (right plot). (The residual 
background under the signal is due to the experimental $E_T^{\rm{miss}}$ resolution.) 
   \item[(4)~~] We also apply a combination of other cuts on: the invariant mass and the azimuthal opening angle of the $\tau b$-jet 
system, where $b$-jet is here the remaining one after 
the reconstruction of the top quark ($m _{\tau b-{\rm{jet}}} >$ 100 GeV and $\Delta\Phi(\tau-{\rm {jet}},b-{\rm{jet}})>1.25$ radians); 
the invariant mass 
of the 
$b\bar{b}$ pair ($m_{bb-{\rm{jet}}}>225$ GeV) and the 
transverse mass of the $\tau b$-jet system ($p_T^{\tau b-{\rm{jet}}}>190$ GeV). 
The cumulative effect of these cuts is the reduction of the $W^\pm$+jets background by 
more than one order of magnitude, while the signal 
($M_{H^\pm} =$ 170~GeV) and the $t\bar{t}$ background are 
suppressed by only a factor of two.
   \item[(5)~~] Finally, we require $m_T > 100$ GeV for the calculation of the 
signal-to-background ratios and the signal significances in Tab.~\ref{tab:table1}. 
This cut is very 
efficient against the $t\bar{t}$ noise (the efficiency is 0.06\% for a 
$M_{H^\pm} =170$ GeV Higgs signal, $1.9\,\times
\,10^{-3}$ and $0.42\,\times\,10^{-6}$ for the $t\bar{t}$ and the $W^\pm$+jets backgrounds, respectively).
\begin{table*}
\begin{center}
\begin{minipage}{.75\linewidth} 
\caption{\label{tab:table1} Sensitivity of the ATLAS detector to the observation of 
charged Higgs bosons through $H^\pm\rightarrow \tau\nu$ decays in the transition region, 
for an integrated luminosity of 30~fb$^{-1}$ and $\tan\beta=50$.}
\end{minipage}
\vspace*{0.25cm}
\vbox{\offinterlineskip 
\halign{&#& \strut\quad#\hfil\quad\cr  
%\colrule
& $M_{H^\pm}$ (GeV)                && 160  && 170  && 180   && 190  & \cr
%\colrule
& Signal ($S$)                     && 35   && 46   &&  50   &&  35  & \cr
& Backgrounds ($B$)                && 13   && 13   &&  13   &&  13  & \cr
& $S/B$                            && 2.7  && 3.5  &&  3.8  &&  2.7 & \cr
& $S/\sqrt{B}$                     && 9.7  && 12.8 &&  13.9 && 9.7  & \cr
& Poisson Significance             && 7.3  && 9.1  &&  9.8  && 7.3  & \cr
& Poisson Significance$+$5\% syst. && 7.1  && 8.9  &&  9.5  && 7.1  & \cr
%\colrule
}}
\end{center}
\end{table*}
\end{description}

\begin{figure}[!htbp]
\begin{center}
\epsfig{file=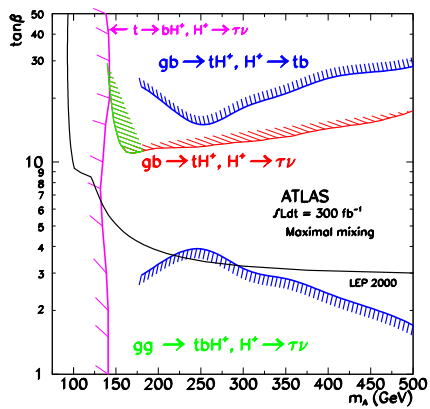,height=6cm}
\caption{\small The new ATLAS discovery potential for charged Higgs bosons. The results
of the current analysis are shown in green.}
\label{fig:newcon}
\vspace{-0.25cm}
\end{center}
\end{figure}

\section{RESULTS}
The discovery contour in the transition region resulting from this new analysis
is shown in Fig.~\ref{fig:newcon}. Notice that, at lower 
masses, the signal reconstruction efficiency decreases (although the rate is higher),
thus explaining the upward turn of the discovery reach.  

Before closing, some additional information is in order regarding the
interplay between the new curve and the two old ones. In fact, recall that
above the top-quark mass, the $2\to 2$ process, $bg \to tH^-$, with $H^\pm \to\tau\nu$, 
was used while below it the charged Higgs was searched for in top-quark decays, 
$t \to bH^\pm$, counting the excess of $\tau$-leptons over the SM expectations. 
Furthermore, in the analysis above the top-quark mass, {\small CTEQ2L} 
PDFs~\cite{PDFs} were used and the charged Higgs production cross sections were obtained from
another generator, 
{\small PYTHIA} v5.7. These differences complicate the matching of the various contours at 
their boundaries, especially between the transition region and the high mass region 
($M_{H^\pm} > m_t$). In the result shown, the normalisation cross sections for the transition 
region were matched to the {\small PYTHIA} v5.7 numbers above $m_t$, for consistency with the previous analysis 
of the high mass region~\cite{Assamagan:2002ne}. A second stage of this analysis is currently underway to 
update all the discovery contours by adopting the same $2\to3$ production process throughout.

\section{CONCLUSIONS}
Meanwhile, as {\sl ad interim} conclusion, we would like to claim that the LHC
discovery potential of charged Higgs bosons has been extended further by our preliminary analysis. 

\section*{ACKNOWLEDGEMENTS}

We would like to thank the 2003 Les Houches workshop organisers
for their kind invitation. SM is grateful to The Royal Society (London, UK)
for a Conference Grant.

%\bibliography{threshold}

\end{document}